\documentstyle[twocolumn,aps]{revtex}
\begin{document}
\draft
\title{\bf {Comment on ``Friedel phases and phases of transmission
amplitudes in quantum scattering systems'' by T. Taniguchi and
M. B{\"u}ttiker}}
\author{P. Singha Deo\cite{eml}} 
\address{Department of Physics, University of Jyv{\"a}skyl{\"a},
P.O.Box 35, 40351 Jyv{\"a}skyl{\"a}, Finland.}
\maketitle
\begin{abstract}
We take a modified boundary condition at the dead end
of a stub to simulate transmission zeroes being replaced
by minima and then the 
discontinuous phase slip (or decrease)
at the transmission
zeroes are replaced by a continuous but rapid phase
slip. The modified
boundary condition can be continuously tuned to give
the results of the stub with hard wall boundary condition
at the dead end of stub. Even when the phase slip
is continuous one can obtain information
about the density of states in the stub region
from the scattering phases.
\end{abstract}
\narrowtext

Transport across the stub structure has acquired a lot of importance
recently \cite{tan}. The stub structure can exhibit transmission zeroes at
certain Fermi energies \cite{bay} and there
is a discontinuous slip (or decrease)
in the phase of the transmission amplitude
as the Fermi energy crosses this transmission zero \cite{deo}. Recently it
was shown \cite{tan} that if one subtracts
this discontinuous phase slip from the phase of
the transmission amplitude then the remaining phase
(which they term Friedel phase) satisfy the Friedel sum rule \cite{tan}.
All these analysis are
based on calculations with a hard wall boundary condition
(an infinite barrier potential or an infinite well potential)
at the dead end of the stub. However, in realistic situations
one may not find absolutely discontinuous phase slip over
zero energy scale. In this comment we find a simple way
to simulate the fact that this phase slip can be continuous
but rapid.
An infinite potential well at the dead end of the stub
reflects an incident electron with unit probability.
Now a small perturbation from this would be a finite but very
deep potential well at the dead end of the stub. This simulates
the fact that electrons are almost entirely reflected from the
end of the stub (a negligible amount escapes), 
transmission zeroes are replaced by minima
(a similar situation was pointed out to me by A. M. Jayannavar
\cite{jay})
and here we show that
the discontinuous phase slip at the transmission zero
is then replaced by a continuous but rapid slip in phase.

In Fig.~1 we show the transmission coefficient $T$ versus
$kL$ across the modified stub
whose length $L$ is normalized to one by choice of the unit of length,
by solid curve. The dotted curve gives the phase of the
transmission amplitude $\theta_t$ . Here $k$ is incident wave vector. 
The transmission
coefficient does not exhibit zeroes and the phase of the
transmission amplitude rapidly decreases at the minima.
However we show that
the idea put forward by Taniguchi and B{\"u}ttiker
in Ref. \cite{tan} can
be extended to get information about density of states
from the scattering phase and in the limit when the
potential well at the end of the stub becomes infinite one
can recover their results.  For
this it is not necessary to define a Friedel phase but one
has to take into account the fact that for hard wall boundary
condition at the dead end of the stub we can write $d\theta_r /d(kL)
=d\theta_t /d(kL)=\rho = \Sigma |\psi_ \mu |^2$ where $\psi_\mu$
is the wavefunction inside the stub and $\theta_r$
is the phase of reflection amplitude. This directly follows
from Friedel sum rule.
In the inset we show that when
$d\theta_t /d(kL)$ deviates from $\rho$ for the modified
stub then
$d\theta_r /d(kL)$ coincides with it and vice versa. The
Gaussian deviations from Friedel sum rule becomes narrower
as the potential at the end of the stub
becomes deeper. The width of these Gaussian deviations
from $\rho$ give the scale over which the continuous
phase slip occur. In the limit of
an infinite well potential these deviations become
delta function like as found in Ref. \cite{tan} and shown
in Fig.~2.

The purpose of this comment is that there can be situations
when the discontinuous phase slip produced by geometric
scattering can be continuous and even then there is
a way to extend the idea of Taniguchi and B{\"u}ttiker \cite{tan}
although in that case defining a Friedel phase is not easy.
It is possible that such situations can be simulated in other
ways like a magnetic field in the quasi one dimensional case.
General violation of Friedel sum rule, when we consider
the scattering phases in all directions is shown elsewhere 
\cite{man}.

\centerline{Figure captions}

\noindent Fig.~1 The solid curve is transmission coefficient T across
a modified
stub. The dotted curve is the phase of the transmission amplitude
$\theta_t$ across the stub. 
At the dead end of the stub there is a potential
well of $VL^2=-100$.
Next we make $VL^2=-10^6$ and in the
inset we plot $d\theta_t/d(kL)$ (dotted curve), $d\theta_r/d(kL)$
(dashed curve) and $\rho$ (solid curve). We have chosen $\hbar=2m=1$.

\noindent Fig.~2 The same as in the inset of Fig.~1 but now
$VL^2=-10^{10}$.

\begin{thebibliography}{99}
\bibitem[*]{eml} Electronic mail: deo@phys.jyu.fi
\bibitem{tan} T. Taniguchi and M. B\"uttiker, Phys. Rev. B {\bf 60},
13814 (1999), and references therein.
\bibitem{bay} B. F. Bayman and C. J. Mehoke, Am. Journ. of Phys.
{\bf 51} 875(1983); W. Porod, Z. Shao and C. S. Lent, Phys. Rev. B
{\bf 48}, 8495(1993) and references therein.
\bibitem{deo} P. Singha Deo, Phys. Rev. B {\bf 53}, 15447 (1996).
\bibitem{jay} A. M. Jayannavar, private communication (1994).
\bibitem{man} P. Singha Deo et al (under preparation).
\end{thebibliography}
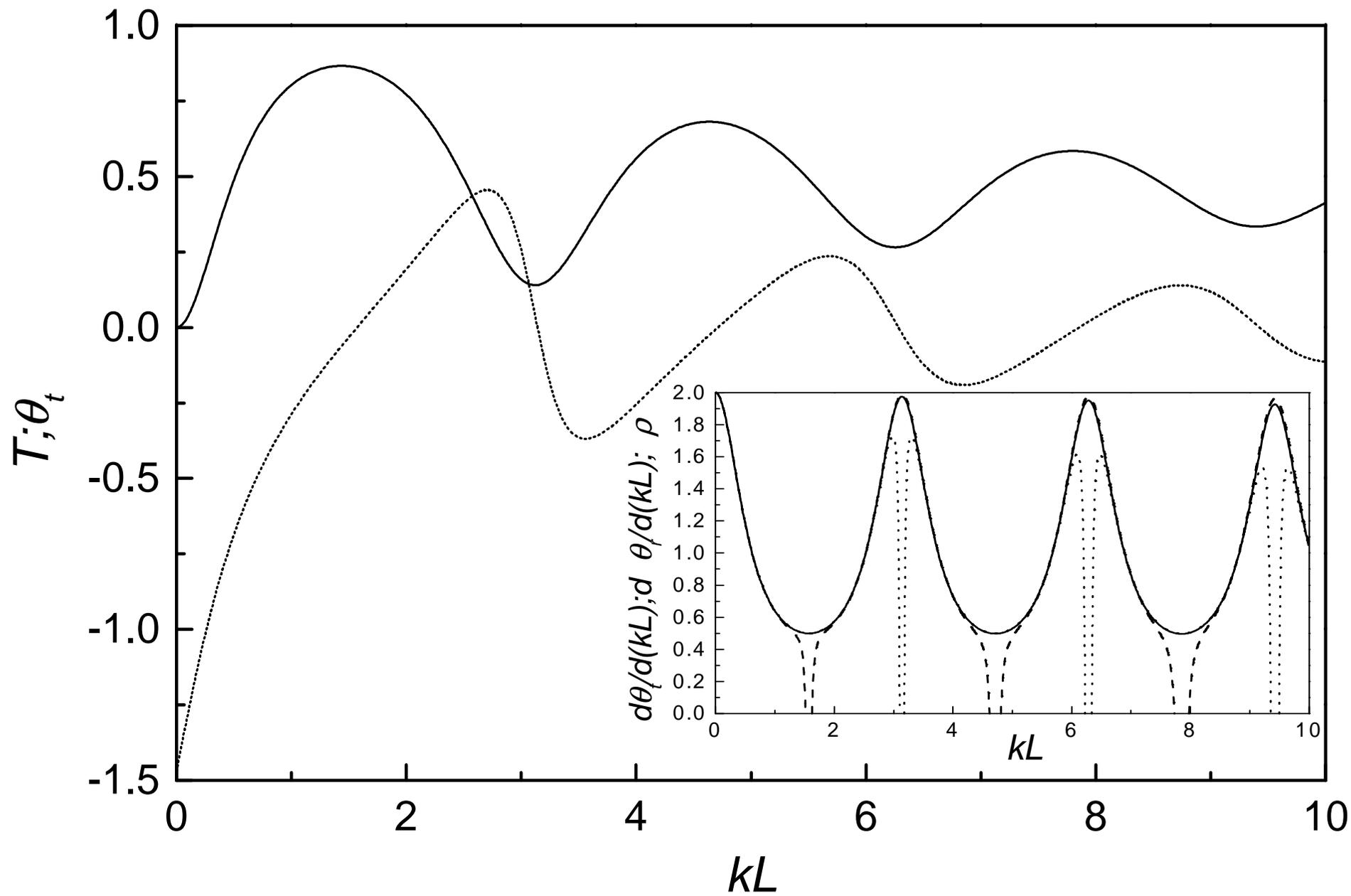
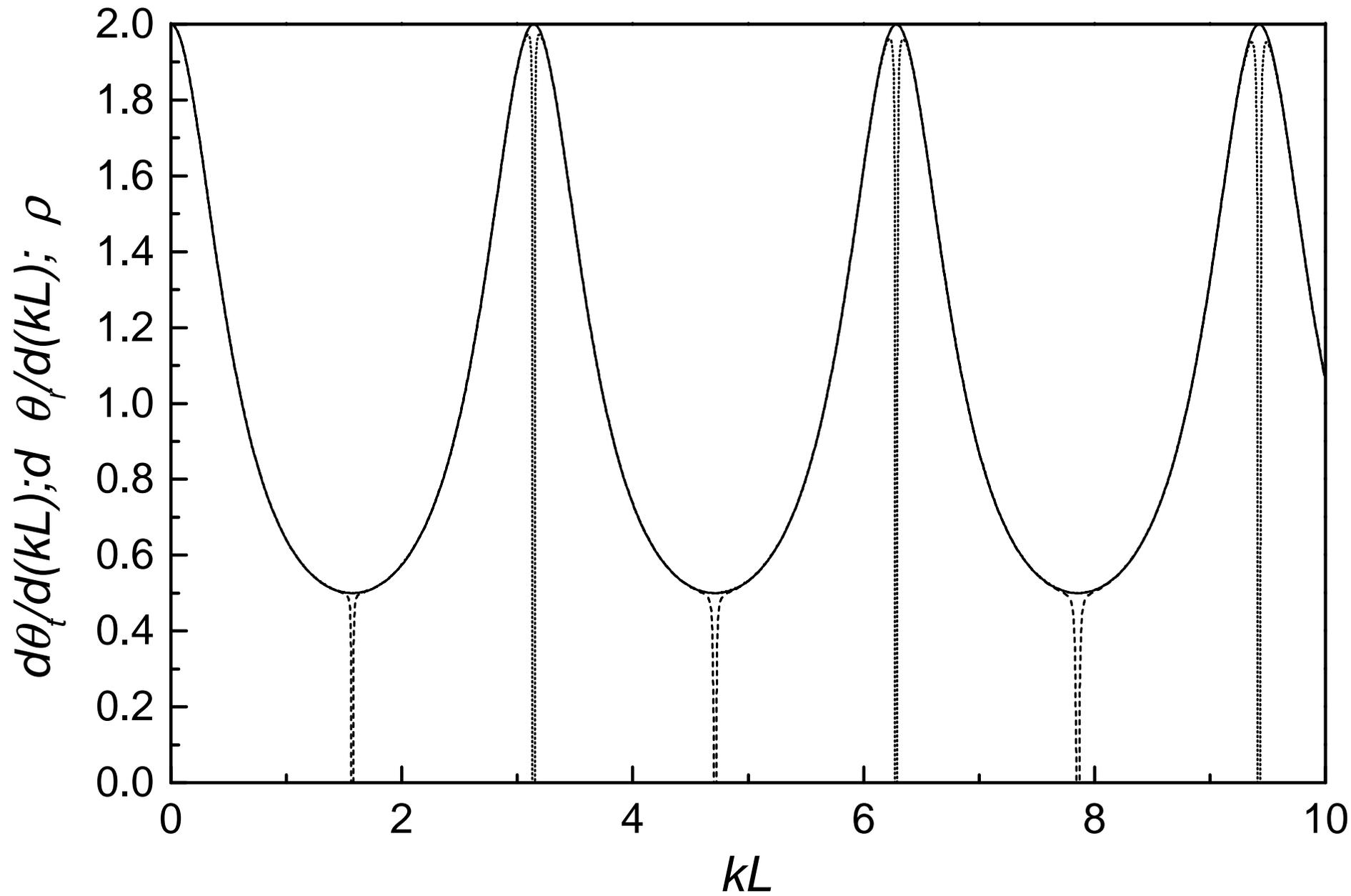
\end{document}